\documentclass[12pt]{article}
\font\twelveof=msym10 at 12pt
\def\Z{\mbox{\twelveof Z}}

\def\C{\mbox{$C_{\lambda}$}}
\def\ap{a^{\dagger}}
\def\alg{${\cal A}^{(\lambda)}_{\alpha_0 \alpha_1 \ldots \alpha_{\lambda-2}}$}
\def\algtwo{${\cal A}^{(2)}_{\alpha_0}$}
\def\algthree{${\cal A}^{(3)}_{\alpha_0 \alpha_1}$}
\def\case#1#2{{\textstyle{#1\over #2}}}
\def\RE{\mathop{\Re e}\nolimits}
\def\IM{\mathop{\Im m}\nolimits}
\def\Qp{Q^{\dagger}}
\def\H{\mbox{$\cal H$}}

\newcommand{\levelone}{\thicklines\line(1,0){10}}
\newcommand{\leveltwo}{\thicklines\line(1,0){8}}
\newcommand{\trait}{\line(0,1){2}}
\newcommand{\ord}{\thicklines\line(1,0){2}}
\newcommand{\separation}{\line(0,1){115}}
\newsavebox{\three}
\newsavebox{\four}
\newsavebox{\dash}
\newsavebox{\Qzero}
\newsavebox{\Qone}
\newsavebox{\Qtwo}

\title{
\hfill{\normalsize ULB/229/CQ/97/5}\\
\vspace{1cm}
$C_{\lambda}$-extended harmonic oscillator and (para)supersymmetric quantum
mechanics}
\author{C. Quesne \thanks{Directeur de recherches FNRS. E-mail:
cquesne@ulb.ac.be}\ , N. Vansteenkiste \thanks{Boursier ULB. E:mail:
nvsteen@ulb.ac.be}\\
{\small \sl Physique Nucl\'eaire Th\'eorique et Physique
Math\'ematique,  Universit\'e Libre de Bruxelles,} \\
{\small \sl Campus de la
Plaine CP229, Boulevard~du Triomphe, B-1050 Brussels, Belgium}}
\date{ }
\begin{document}
\baselineskip=22pt plus 1pt minus 1pt
%%%%%%%%%%%%%%%%%%%%%%%%%%%%%%%%%%%%%%%%%%%%%%%%%%%%%%%%%%
\maketitle

\begin{abstract}
$\C$-extended oscillator algebras are realized as generalized deformed
oscillator
algebras. For $\lambda = 3$, the spectrum of the corresponding bosonic
oscillator
Hamiltonian is shown to strongly depend on the algebra parameters. A connection
with cyclic shape invariant potentials is noted. A bosonization of PSSQM of
order
two is obtained.
\end{abstract}

\vspace{1.5cm}



\vspace{1.5cm}

\noindent
To be published in Phys. Lett. A
\newpage
%
%========================================================================
%
\section{Introduction}
During the last few years, exotic quantum statistics have received considerable
attention in the literature. In two spatial dimensions, one can have anyonic
statistics~\cite{leinaas}, interpolating between bosonic and fermionic
ones, which
has been proposed as a mechanism for the fractional quantum Hall effect and
high-$T_c$ superconductivity~\cite{wilczek}. In higher dimensions, parabosonic
and parafermionic statistics have been suggested as generalizations of the
standard bosonic and fermionic ones, describing representations of the
permutation
group that are neither completely symmetrical nor completely
antisymmetrical~\cite{green}.\par
%
%------------------------------------------------------------------------
%
On the other hand, supersymmetric quantum mechanics~(SSQM) has established a
nice symmetry between bosons and fermions~\cite{witten}. Furthermore, when
supplemented with the concept of shape invariance~\cite{gendenshtein}, it
has also
provided a powerful method of generating exactly solvable quantum mechanical
models. Devising new approaches to construct shape invariant potentials is still
under current investigation (for a recent review see Ref.~\cite{cooper}). Among the
most recent advances in this field, one may quote the introduction of
cyclic shape
invariant potentials by Sukhatme {\it et al.}~\cite{sukhatme}, generalizing a
previous work of Gangopadhyaya and Sukhatme~\cite{gango}.\par
%
%------------------------------------------------------------------------
%
In view of the SSQM success, there have been various attempts to extend its
formalism to some of the exotic statistics. Combining for instance bosons with
parafermions (instead of fermions) has led to parasupersymmetric quantum
mechanics~(PSSQM), existing in the literature in two (generally) inequivalent
forms, due to Rubakov-Spiridonov~\cite{rubakov}, and
Beckers-Debergh~\cite{beckers90}, respectively.\par
%
%-----------------------------------------------------------------------
%
The development of quantum groups and quantum algebras~\cite{drinfeld} during
the last decade has proved very useful in connection with such problems. Various
deformations and extensions of the oscillator algebra (for a recent review see
Ref.~\cite{katriel}) have indeed been applied to the description of systems with
nonstandard statistics~\cite{greenberg, chaturvedi, cq94, meljanac, plyu}, the
algebraic formulation of some quantum integrable models~\cite{vasiliev, poly,
brze}, and the bosonization of SSQM~\cite{plyu, brze, bonatsos}.\par
%
%-------------------------------------------------------------------------
%
The purpose of the present letter is to introduce new $\C$-extended oscillator
algebras, where $C_{\lambda}$ denotes a cyclic group of order~$\lambda$, and to
show their usefulness in both SSQM and PSSQM frameworks.\par
%
%========================================================================
%
\section{\C-extended oscillator algebras}
A \C-extended oscillator algebra ${\cal A}^{(\lambda)}$, where $\lambda$ may
take any value in the set $\{\,2, 3, 4, \ldots\,\}$, is defined as an
algebra generated
by the operators $I$, $\ap$ , $a = \left(\ap\right)^{\dagger}$, $N =
N^{\dagger}$, and
$T = \left(T^{\dagger}\right)^{-1}$, satisfying the relations
\begin{eqnarray}
  \left[N, \ap\right] & = & \ap, \qquad [N, T] = 0, \qquad T^{\lambda} = I,
\nonumber
            \\
  \left[a, \ap\right] & = & I + \sum_{\mu=1}^{\lambda-1} \kappa_{\mu} T^{\mu},
            \qquad \ap T = e^{-2\pi i/\lambda}\, T \ap,   \label{eq:alg-def}
\end{eqnarray}
together with their Hermitian conjugates. Here $\kappa_{\mu}$, $\mu = 1$,
2, $\ldots$,~$\lambda-1$, are some complex parameters restricted by the
conditions $\kappa_{\mu}^* = \kappa_{\lambda - \mu}$ (so that there remain
altogether $\lambda-1$ independent real parameters), and $T$ is the
generator of a
cyclic group of order~$\lambda$, $\C = \{\,I, T, T^2, \ldots,
T^{\lambda-1}\,\}$ (or,
more precisely, the generator of a unitary representation thereof). As
usual, $N$,
$\ap$, and $a$ will be called number, creation, and annihilation operators,
respectively.\par
%
%-------------------------------------------------------------------------
%
As well known~\cite{cornwell}, \C\ has $\lambda$ inequivalent unitary
irreducible
matrix representations~$\Gamma^{\mu}$, $\mu = 0$, 1, $\ldots$,~$\lambda-1$,
which are one-dimensional, and such that $\Gamma^{\mu}\left(T^{\nu}\right) =
\exp(2\pi i \mu \nu/\lambda)$ for any $\nu = 0$, 1, $\ldots$,~$\lambda-1$. The
projection operator on the carrier space of~$\Gamma^{\mu}$ may be written as
\begin{equation}
  P_{\mu} = \frac{1}{\lambda} \sum_{\nu=0}^{\lambda-1} \Bigl(\Gamma^{\mu} \left(
  T^{\nu}\right) \Bigr)^* T^{\nu} = \frac{1}{\lambda} \sum_{\nu=0}^{\lambda-1}
  e^{-2\pi i\mu\nu/\lambda}\, T^{\nu},   \label{eq:proj}
\end{equation}
and conversely $T^{\nu}$, $\nu=0$, 1, $\ldots$,~$\lambda-1$, may be expressed in
terms of the $P_{\mu}$'s as
\begin{equation}
  T^{\nu} = \sum_{\mu=0}^{\lambda-1}  e^{2\pi i\mu\nu/\lambda} P_{\mu}.
\end{equation}
\par
%
%-------------------------------------------------------------------------
%
The algebra defining relations~(\ref{eq:alg-def}) may therefore be rewritten in
terms of $I$, $\ap$, $a$, $N$, and~$P_{\mu}^{\vphantom{\dagger}} =
P_{\mu}^{\dagger}$, $\mu=0$, 1, $\ldots$,~$\lambda-1$, as
\begin{eqnarray}
  \left[N, \ap\right] & = & \ap, \qquad \left[N, P_{\mu}\right] = 0, \qquad
            \sum_{\mu=0}^{\lambda-1} P_{\mu} = I, \nonumber \\
  \left[a, \ap\right] & = & I + \sum_{\mu=0}^{\lambda-1} \alpha_{\mu} P_{\mu},
            \qquad \ap P_{\mu} = P_{\mu+1}\, \ap,   \label{eq:alg-def-bis}
\end{eqnarray}
where we use the conventions $P_{\lambda} \equiv P_0$, $P_{-1} \equiv
P_{\lambda-1}$ (and similarly for other operators or parameters indexed
by~$\mu$). By definition of projection operators, the $P_{\mu}$'s satisfy the
relations $P_{\mu} P_{\nu} = \delta_{\mu,\nu} P_{\mu}$.
Equation~(\ref{eq:alg-def-bis}) depends upon $\lambda$ real parameters
$\alpha_{\mu} = \sum_{\nu=1}^{\lambda-1} \exp(2\pi i\mu\nu/\lambda)
\kappa_{\nu}$, $\mu=0$, 1, $\ldots$,~$\lambda-1$, restricted by the condition
$\sum_{\mu=0}^{\lambda-1} \alpha_{\mu} = 0$. Hence, we may eliminate one of
them, for instance $\alpha_{\lambda-1}$, and denote the algebra by \alg. It will
however often prove convenient to work instead with the $\lambda$ dependent
parameters $\alpha_0$, $\alpha_1$, $\ldots$,~$\alpha_{\lambda-1}$.\par
%
%------------------------------------------------------------------------
%
We may realize \C\ in various ways. Two of the simplest ones use either
functions
of~$N$, or functions of spin-$s$ matrices, where $s \equiv (\lambda-1)/2$. Here,
we will consider the former choice. Hence, in the remainder of this letter,
we will
assume that $T$ and (as a consequence of Eq.~(\ref{eq:proj})) $P_{\mu}$ are
given
by
\begin{equation}
  T = e^{2\pi iN/\lambda}, \qquad P_{\mu} = \frac{1}{\lambda}
  \sum_{\nu=0}^{\lambda-1} e^{2\pi i \nu (N-\mu)/\lambda}, \qquad \mu = 0, 1,
  \ldots, \lambda-1,    \label{eq:N-realize}
\end{equation}
respectively.\par
%
%------------------------------------------------------------------------
%
With such a choice, \alg\ may be considered as a generalized deformed oscillator
algebra (GDOA) ${\cal A}(G(N))$, with $G(N) = I + \sum_{\mu=0}^{\lambda-1}
\alpha_{\mu} P_{\mu}$, and $P_{\mu}$ given by
Eq.~(\ref{eq:N-realize})~\cite{cq95}. For any GDOA, one may define a so-called
structure function~$F(N)$, which is the solution of the difference equation
$F(N+1)
- F(N) = G(N)$, such that $F(0) = 0$ (see Ref.~\cite{katriel} and
references quoted
therein). In the present case, a straightforward calculation leads to
\begin{equation}
  F(N) = N + \sum_{\mu=0}^{\lambda-1} \beta_{\mu} P_{\mu}, \qquad \beta_0
  \equiv 0, \qquad \beta_{\mu} \equiv \sum_{\nu=0}^{\mu-1} \alpha_{\nu} \quad
  (\mu =1, 2, \ldots, \lambda-1).   \label{eq:F}
\end{equation}
\par
%
%------------------------------------------------------------------------
%
It has been shown~\cite{cq96} that GDOAs may have in general various types of
unitary irreducible representations according the nature of the
$N$~spectrum, but
here we shall only be interested in the bosonic Fock-space representation,
wherein
\begin{equation}
  \ap a = F(N), \qquad a \ap = F(N+1).   \label{eq:F-Fock}
\end{equation}
Its carrier space is spanned by the eigenvectors~$|n\rangle$ of the number
operator~$N$, corresponding to the eigenvalues $n=0$, 1, 2,~$\ldots$, where
$|0\rangle$ is assumed to be a vacuum state, i.e., $a |0\rangle = 0$. The
eigenvectors can be written as
\begin{equation}
  |n\rangle = {\cal N}_n^{-1/2} \left(\ap\right)^n |0\rangle, \qquad n = 0,
1, 2,
  \ldots,    \label{eq:vectors}
\end{equation}
where ${\cal N}_n = \prod_{i=1}^n F(i)$. By writing $n$ as $n = k\lambda + \mu$,
where $\mu \in \{\,0, 1, \ldots, \lambda-1\,\}$, and $k$ is some nonnegative
integer, ${\cal N}_n$ can be expressed in terms of gamma functions as
\begin{equation}
  {\cal N}_{k\lambda+\mu} = \lambda^{k\lambda+\mu} \left(\prod_{\nu=0}^{\mu}
  \Gamma(k+1+ \overline{\beta}_{\nu})\right)
\left(\prod_{\nu'=\mu+1}^{\lambda-1}
  \Gamma(k+ \overline{\beta}_{\nu'})\right) \left(\prod_{\nu"=1}^{\lambda-1}
  \Gamma(\overline{\beta}_{\nu"})\right)^{-1},   \label{eq:norm}
\end{equation}
where $\overline{\beta}_{\nu} \equiv \left(\beta_{\nu} +
\nu\right)/\lambda$. The
creation and annihilation operators act upon~$|n\rangle$ in the usual way, i.e.,
\begin{equation}
  \ap |n\rangle = \sqrt{F(n+1)}\, |n+1\rangle, \qquad a |n\rangle =
\sqrt{F(n)}\,
  |n-1\rangle,
\end{equation}
while $P_{\mu}$ projects on the $\mu$th component ${\cal F}_{\mu} \equiv
\{\, |k\lambda + \mu\rangle \mid k = 0, 1, 2, \ldots\,\}$ of the
\Z$_{\lambda}$-graded Fock space ${\cal F} = \sum_{\mu=0}^{\lambda-1} \oplus
{\cal F}_{\mu}$. It is obvious that such a Fock-space representation exists
if and
only if $F(\mu) > 0$ for $\mu=1$, 2, $\ldots$,~$\lambda-1$. These conditions
imply the following restrictions on the parameters~$\alpha_{\mu}$,
\begin{equation}
  \sum_{\nu=0}^{\mu-1} \alpha_{\nu} > - \mu, \qquad \mu = 1, 2, \ldots,
\lambda-1.
  \label{eq:cond-Fock}
\end{equation}
\par
%
%-------------------------------------------------------------------------
%
{}For the lowest allowed $\lambda$ value, i.e., $\lambda=2$, the operator~$T$,
defined in Eq.~(\ref{eq:N-realize}), reduces to the Klein operator $K =
\exp(i\pi N)$,
which in the Fock-space representation becomes $K=(-1)^N$. The corresponding
operators $P_0 = \frac{1}{2} \left(I + (-1)^N\right)$, and $P_1 =
\frac{1}{2} \left(I
- (-1)^N\right)$ project upon the even subspace ${\cal F}_0 = \{\,
|2k\rangle \mid
k = 0, 1, 2, \ldots\,\}$, and the odd subspace ${\cal F}_1 = \{\,
|2k+1\rangle \mid
k = 0, 1, 2, \ldots\,\}$ of $\cal F$, respectively. The $C_2$-extended
oscillator
algebra \algtwo\ is therefore nothing else than the Calogero-Vasiliev oscillator
algebra~\cite{vasiliev}, providing an algebraic formulation of the two-particle
Calogero problem~\cite{poly, brze}, an alternative description of
parabosons~\cite{chaturvedi}, and a bosonization of SSQM~\cite{plyu, brze}. It
depends upon a single independent real parameter $\kappa_1 = \kappa_1^* =
\alpha_0 = - \alpha_1$, restricted by the condition $\alpha_0 > -1$ in the
Fock-space representation.\par
%
%------------------------------------------------------------------------
%
The next allowed $\lambda$ value, i.e., $\lambda=3$, gives rise to a new
algebraic
structure, the $C_3$-extended oscillator algebra \algthree, corresponding
to $T =
\exp(2\pi i N/3)$, and
\begin{eqnarray}
  P_0 & = & \case{1}{3} \left(I + 2 \cos\case{2\pi}{3}N\right), \qquad
           P_1 = \case{1}{3} \left(I - \cos\case{2\pi}{3}N + \sqrt{3}
           \sin\case{2\pi}{3}N \right), \nonumber \\
  P_2 & = & \case{1}{3} \left(I - \cos\case{2\pi}{3}N - \sqrt{3}
\sin\case{2\pi}{3}N
           \right),
\end{eqnarray}
projecting on ${\cal F}_0 = \{\, |3k\rangle \mid k = 0, 1, 2, \ldots\,\}$,
${\cal F}_1
= \{\, |3k+1\rangle \mid k = 0, 1, 2, \ldots\,\}$, and ${\cal F}_2 = \{\,
|3k+2\rangle
\mid k = 0, 1, 2, \ldots\,\}$, respectively. It depends upon two
independent real
parameters, which may be taken as the real and imaginary parts of~$\kappa_1$
(with $\kappa_2 = \kappa_1^*$), or as $\alpha_0 = 2 \RE\kappa_1$, and $\alpha_1
= - \RE\kappa_1 - \sqrt{3} \IM\kappa_1$ (with $\alpha_2 = - \alpha_0 - \alpha_1
= - \RE\kappa_1 + \sqrt{3} \IM\kappa_1$), restricted by the conditions $\alpha_0
> -1$, and $\alpha_0 + \alpha_1 > -2$ in the Fock-space representation. For such
an algebra, we may write
\begin{eqnarray}
  \left[a, \ap\right] & = & I + 2 (\RE\kappa_1) \cos\case{2\pi}{3}N
            - 2 (\IM\kappa_1) \sin\case{2\pi}{3}N \nonumber \\
  & = & I + \alpha_0 P_0 + \alpha_1 P_1 - (\alpha_0 + \alpha_1) P_2.
\end{eqnarray}
\par
%
%------------------------------------------------------------------------
%
Similar explicit relations can easily be written down for $\lambda = 4$,
5,~$\ldots$.\par
%
%========================================================================
%
\section{\C-extended oscillator Hamiltonian and supersymmetric quantum
mechanics}
As usual, we define the bosonic oscillator Hamiltonian associated to the algebra
\alg, in appropriate units, as
\begin{equation}
  H_0 \equiv \case{1}{2} \left\{a, \ap\right\}.   \label{eq:Hzero}
\end{equation}
By using Eqs.~(\ref{eq:alg-def-bis}), ({\ref{eq:F}), and~(\ref{eq:F-Fock}),
$H_0$ can
be rewritten in the equivalent forms
\begin{equation}
  H_0 = \ap a + \frac{1}{2} \left(I + \sum_{\mu=0}^{\lambda-1} \alpha_{\mu}
  P_{\mu}\right) = N + \frac{1}{2} I + \sum_{\mu=0}^{\lambda-1} \gamma_{\mu}
  P_{\mu},
\end{equation}
where the parameters~$\gamma_{\mu}$ are defined by
\begin{equation}
  \gamma_{\mu} \equiv \case{1}{2} (\beta_{\mu} + \beta_{\mu+1}) =
  \left\{\begin{array}{ll}
        \case{1}{2} \alpha_0 & \mbox{if $\mu=0$}, \\[0.2cm]
        \sum_{\nu=0}^{\mu-1} \alpha_{\nu} + \case{1}{2} \alpha_{\mu} & \mbox{if
              $\mu=1, 2, \ldots, \lambda-1$},
  \end{array}\right.   \label{eq:gamma}
\end{equation}
and satisfy the relation $\sum_{\mu=0}^{\lambda-1}\, (-1)^{\mu} \gamma_{\mu} =
0$.\par
%
%------------------------------------------------------------------------
%
The eigenvectors of $H_0$ are the states~$|n\rangle$, defined in
Eqs.~(\ref{eq:vectors}) and~(\ref{eq:norm}), and their eigenvalues are given by
\begin{equation}
  E_{k\lambda+\mu} = k\lambda + \mu + \gamma_{\mu} + \case{1}{2}, \qquad k = 0,
 1, 2, \ldots, \qquad \mu = 0, 1, \ldots, \lambda-1.
\end{equation}
In each ${\cal F}_{\mu}$ subspace of the $\Z_{\lambda}$-graded Fock space~$\cal
F$, the spectrum of~$H_0$ is therefore harmonic, but the $\lambda$ infinite
sets of
equally spaced energy levels, corresponding to $\mu=0$, 1,
$\ldots$,~$\lambda-1$, may be shifted with respect to each other by some
amounts depending upon the algebra parameters $\alpha_0$, $\alpha_1$,
$\ldots$,~$\alpha_{\lambda-2}$, through their linear combinations
$\gamma_{\mu}$, $\mu=0$, 1, $\ldots$,~$\lambda-1$.\par
%
%-----------------------------------------------------------------------
%
{}For the Calogero-Vasiliev oscillator, the two infinite sets of energy levels
corresponding to ${\cal F}_0$ and ${\cal F}_1$, respectively, are always shifted
with respect to one another by one energy unit, since the relation $\gamma_0 =
\gamma_1$ is valid for any $\alpha_0$~value. The resulting spectrum is therefore
very simple, and coincides with that of a shifted harmonic oscillator.\par
%
%------------------------------------------------------------------------
%
{}For $\lambda\ge 3$, the situation is entirely different. According to the
parameter
values, the spectrum may be nondegenerate, or may exhibit some ($\nu+1$)-fold
degeneracies above some energy eigenvalue, where $\nu=1$, 2,
$\ldots$,~$\lambda-2$, or~$\lambda-1$. Already for the $\lambda=3$ case, which
we did fully analyze, one gets a lot of different types of spectra. Here,
we will
merely sketch their classification and provide some examples, leaving the full
discussion for a forthcoming publication~\cite{cq97a}.\par
%
%-----------------------------------------------------------------------
%
Starting with the nondegenerate case, it can easily be shown that the
ground states
in ${\cal F}_0$, ${\cal F}_1$, and ${\cal F}_2$ may be ordered in three
different
ways, which we will refer to as I, II, and~III, respectively, as listed
hereafter
\begin{equation}
  \begin{array}{lll}
     \mbox{(I)} & E_0 < E_1 < E_2 \quad & \mbox{if $-1 < \alpha_0 < 2$ and
              $-2-\alpha_0 < \alpha_1$}, \\[0.2cm]
     \mbox{(II)} & E_0 < E_2 < E_1 \quad & \mbox{if $2 < \alpha_0$ and
              $-4 < \alpha_1$}, \\[0.2cm]
     \mbox{(III)} & E_2 < E_0 < E_1 \quad & \mbox{if $2 < \alpha_0$ and
              $-2-\alpha_0 < \alpha_1 < -4$}.
  \end{array}
\end{equation}
Comparing now the positions of the excited states leads one to divide each class
into subclasses, themselves labelled by one or two integer indices: (I.1.$n$),
(I.2.$n$), (II.1.$m$.$n$), (II.2.$m$.$n$), (III.1.$m$.$n$),
(III.2.$m$.$n$), where $m$,
$n=1$, 2, 3,~$\ldots$. Type~(I.1.$n$) spectra, for instance, correspond to the
following ordering and parameter values
\begin{equation}
  \begin{array}{ll}
    \mbox{(I.1.$n$)} & E_0 < E_3 < \cdots < E_{3n-3} < E_1 < E_2 < E_{3n} <
E_4 < E_5
            < \cdots \\[0.2cm]
    & \mbox{if $-1 < \alpha_0 < 2$ and $6n - \alpha_0 - 8 < \alpha_1 < 6n - 4$}.
  \end{array}
\end{equation}
We note that only for~$n=1$, such an ordering coincides with that of the
standard
harmonic oscillator.\par
%
%------------------------------------------------------------------------
%
Considering next the doubly-degenerate case, one again finds various
possibilities
by reviewing all the limiting cases of the nondegenerate one: (I.$n$.a),
(I.$n$.b),
(II.$m$.$n$.a), (II.$m$.$n$.b), (II.$m$.$n$.c), (III.$m$.$n$.a),
(III.$m$.$n$.b),
(III.$m$.$n$.c), where $m$, $n=1$, 2, 3,~$\ldots$, and a, b, c refer to ${\cal
F}_0$--${\cal F}_1$, ${\cal F}_0$--${\cal F}_2$, and ${\cal F}_1$--${\cal F}_2$
degeneracies, respectively. Type~(I.$n$.a) spectra, for instance, are given by
\begin{equation}
  \begin{array}{ll}
    \mbox{(I.$n$.a)} & E_0 < E_3 < \cdots < E_{3n-3} < E_{3n} = E_1 < E_2 <
E_{3n+3} =
            E_4 < E_5 < \cdots \\[0.2cm]
    & \mbox{if $-1 < \alpha_0 < 2$ and $\alpha_1 = 6n - \alpha_0 - 2$}.
  \end{array}
\end{equation}
\par
%
%------------------------------------------------------------------------
%
The triply-degenerate case is finally dealt with in a similar way by
starting from
the doubly-degenerate one. Here, one only gets three possibilities,
referred to as
(I.$n$.abc), (II.$m$.$n$.abc), (III.$m$.$n$.abc), where $m$, $n=1$, 2,
3,~$\ldots$. For
instance,
\begin{equation}
  \begin{array}{ll}
    \mbox{(I.$n$.abc)} & E_0 < E_3 < \cdots < E_{3n-3} < E_{3n} = E_1 = E_2 <
            E_{3n+3} = E_4 = E_5 < \cdots \\[0.2cm]
    & \mbox{if $\alpha_0 = 2$ and $\alpha_1 = 6n - 4$}.
  \end{array}
\end{equation}
\par
%
%------------------------------------------------------------------------
%
We will now show that some of these spectra occur in SSQM when considering
cyclic shape invariant potentials~\cite{sukhatme}. The Hamiltonians
corresponding
to such potentials have an infinite number of periodically spaced
eigenvalues or, in
other words, the level spacings are given by $\omega_0$, $\omega_1$, $\ldots$,
$\omega_{\lambda-1}$, $\omega_0$, $\omega_1$, $\ldots$, $\omega_{\lambda-1}$,
$\omega_0$, $\omega_1$,~$\ldots$. The ground state energy vanishing, the general
formula for the excited energy levels is $k \Omega_{\lambda} +
\sum_{\nu=0}^{\mu} \omega_{\nu}$, where $k=0$, 1, 2,~$\ldots$, $\mu=0$, 1,
$\ldots$,~$\lambda-1$, and $\Omega_{\lambda} \equiv
\sum_{\mu=0}^{\lambda-1} \omega_{\mu}$.\par
%
%-----------------------------------------------------------------------
%
{}From $\lambda=3$ onwards, the shifted and rescaled Hamiltonian
\begin{equation}
  H'_0 \equiv \frac{\Omega_{\lambda}}{\lambda} \left(H_0 - E_{gs}\right),
  \label{eq:Hprime}
\end{equation}
where $H_0$ is given by Eq.~(\ref{eq:Hzero}), and $E_{gs}$ denotes its
ground state
energy, has the same type of spectrum for some parameter values. For $\lambda=3$
for instance, one finds that the generic case $\omega_0 \ne \omega_1 \ne
\omega_2 \ne \omega_0$ is obtained for the (I.1.1), (II.1.1.1), and
(III.1.1.1) type
spectra, corresponding to the orderings $E_{gs} = E_0 < E_1 < E_2 < E_3 <
E_4 < E_5
< \cdots$, $E_{gs} = E_0 < E_2 < E_1 < E_3 < E_5 < E_4 < \cdots$, $E_{gs} =
E_2 < E_0
< E_1 < E_5 < E_3 < E_4 < \cdots$, and the parameter values ($-1 < \alpha_0
< 2$,
$-2 -\alpha_0 < \alpha_1 < 2$), ($2 < \alpha_0 < 8$, $-4 < \alpha_1 < 4
-\alpha_0$),
($2 < \alpha_0 < 8$, $-2 -\alpha_0 < \alpha_1 < -4$), respectively. An
example for
each type is displayed on Fig.~1.\par
%
%-----------------------------------------------------------------------
%
It is not surprising that for $\lambda=2$, the Hamiltonian $H'_0$ does not
follow
the general rule valid for $\lambda \ge 3$. It is indeed well known to be
equivalent
to the two-particle Calogero Hamiltonian~\cite{poly, brze}, whereas
Gangopadhyaya
and Sukhatme~\cite{gango} established that $\lambda=2$ cyclic shape invariant
potentials include in addition a $\delta$-function singularity at $x=0$.\par
%
%=======================================================================
%
\section{Bosonization of parasupersymmetric quantum mechanics}
{}From the results of the previous section, it is clear that the bosonic
oscillator
Hamiltonian~$H_0$, associated to the algebra \alg, has $\lambda$~series of
levels,
which if properly shifted with respect to one another, can be made to
coincide at
least starting from some excited state. Such a spectrum being reminiscent
of that
of PSSQM Hamiltonians of order $p = \lambda-1$, this hints at a possibility of
describing PSSQM in terms of solely boson-like particles, instead of a
combination of bosons and parafermions of order~$p$, as is usually the
case~\cite{rubakov, beckers90}. In support to this idea, one should keep in mind
that the Calogero-Vasiliev algebra~\algtwo\ provides a bosonization of ordinary
SSQM~\cite{plyu, brze}, which is nothing else than PSSQM of order~one. Here, we
will restrict ourselves to the case of PSSQM of order two, corresponding to the
algebra \algthree, but our results can be generalized to arbitrary order
$p$~\cite{cq97b}.\par
%
%------------------------------------------------------------------------
%
Let us recall that in PSSQM of order two, the parasupercharge operators $Q$,
$\Qp$, and the parasupersymmetric Hamiltonian~$\H$ obey the relations
\begin{equation}
  Q^3 = 0 \quad \mbox{(with $Q^2 \ne 0$)}, \qquad [\H, Q] = 0,
\label{eq:cond1}
\end{equation}
and either
\begin{equation}
  Q^2 \Qp + Q \Qp Q + \Qp Q^2 = 4 Q \H,     \label{eq:cond2a}
\end{equation}
or
\begin{equation}
  \left[Q, \left[\Qp, Q\right]\right] = 2 Q \H,    \label{eq:cond2b}
\end{equation}
according to whether one chooses Rubakov-Spiridonov~\cite{rubakov}, or
Beckers-Debergh~\cite{beckers90} approach. They also satisfy the Hermitian
conjugated relations, which we shall take as understood in the remainder of this
section.\par
%
%------------------------------------------------------------------------
%
As ans\" atze for the operators $Q$ and~$\H$, let us choose
\begin{equation}
  Q = \sum_{\nu=0}^{\lambda-1} \left(\xi_{\nu} a + \eta_{\nu} \ap\right)
P_{\nu},
  \qquad \H = H_0 + \case{1}{2} \sum_{\nu=0}^{\lambda-1} r_{\nu} P_{\nu},
  \label{eq:ansatz}
\end{equation}
where $\xi_{\nu}$, $\eta_{\nu}$ are some complex constants, and $r_{\nu}$ some
real ones, to be selected in such a way that Eqs.~(\ref{eq:cond1}), and
(\ref{eq:cond2a}) [or (\ref{eq:cond2b})] are satisfied.\par
%
%----------------------------------------------------------------------
%
Inserting the expression of~$Q$, given in Eq.~(\ref{eq:ansatz}), into the first
condition in Eq.~(\ref{eq:cond1}), one obtains some restrictions on the
parameters
$\xi_{\nu}$, $\eta_{\nu}$, leading to two sets of three independent solutions
for~$Q$. The solutions belonging to the first set may be distinguished by
an index
$\mu \in \{\,0, 1, 2\,\}$, and are given by
\begin{equation}
  Q_{\mu} = \left(\xi_{\mu+1}\, a + \eta_{\mu+1}\, \ap\right) P_{\mu+1} +
  \eta_{\mu+2}\, \ap P_{\mu+2}, \qquad \eta_{\mu+1}, \eta_{\mu+2} \ne 0,
  \label{eq:Qmu}
\end{equation}
while those belonging to the second set can be obtained from the former by
interchanging the roles of $Q$ and~$\Qp$, and will therefore be omitted.\par
%
%------------------------------------------------------------------------
%
Let us consider next the second condition in Eq.~(\ref{eq:cond1}) with~$Q_{\mu}$
given by Eq.~(\ref{eq:Qmu}) for some $\mu \in \{\,0, 1, 2\,\}$, and with the
corresponding parasupersymmetric Hamiltonian~$\H_{\mu}$ also indexed by~$\mu$.
After some straightforward algebra, one gets the restrictions
\begin{equation}
  \xi_{\mu+1} = 0, \qquad r_{\mu} = - 2 + \alpha_{\mu+1} + r_{\mu+2}, \qquad
  r_{\mu+1} = 2 - \alpha_{\mu} + r_{\mu+2},   \label{eq:restrictions}
\end{equation}
so that at this stage one is left with three arbitrary constants $\eta_{\mu+1}$,
$\eta_{\mu+2}$,~$r_{\mu+2}$, the first two being complex, and the third
real.\par
%
%------------------------------------------------------------------------
%
It now remains to impose the third condition~(\ref{eq:cond2a})
or~(\ref{eq:cond2b}). To simultaneously deal with both possibilities, let us
consider the more general condition
\begin{equation}
  u_{\mu} Q_{\mu}^2 \Qp_{\mu} + v_{\mu} Q_{\mu} \Qp_{\mu} Q_{\mu} + w_{\mu}
  \Qp_{\mu} Q_{\mu}^2 = 4 Q_{\mu} \H_{\mu},   \label{eq:cond2-gen}
\end{equation}
where $u_{\mu}$, $v_{\mu}$,~$w_{\mu}$ are some complex constants.
Equations~(\ref{eq:cond2a}) and~(\ref{eq:cond2b}) correspond to $u_{\mu} =
v_{\mu} = w_{\mu} = 1$, and $u_{\mu} = w_{\mu} = - \frac{1}{2} v_{\mu} = -2$,
respectively.\par
%
%-----------------------------------------------------------------------
%
It can be shown that when taking the previous restrictions
(\ref{eq:Qmu}),~(\ref{eq:restrictions}) into account,
Equation~(\ref{eq:cond2-gen})
can be satisfied for two different choices of $\{\, u_{\mu}, v_{\mu},
w_{\mu} \,\}$
provided some new additional conditions are fulfilled:
\begin{equation}
  u_{\mu} = v_{\mu} = w_{\mu} = \frac{4}{|\eta_{\mu+2}|^2 + |\eta_{\mu+1}|^2}
  \qquad \mbox{if} \qquad r_{\mu+2} = (1 + \alpha_{\mu+2})
  \frac{|\eta_{\mu+2}|^2 - |\eta_{\mu+1}|^2}{|\eta_{\mu+2}|^2 +
  |\eta_{\mu+1}|^2},  \label{eq:sol1}
\end{equation}
or
\begin{eqnarray}
  u_{\mu} & \ne & \frac{4}{|\eta_{\mu+2}|^2 + |\eta_{\mu+1}|^2}, \qquad
v_{\mu} =
           \frac{1}{|\eta_{\mu+2}|^2} \left(4 - |\eta_{\mu+1}|^2 u_{\mu}\right),
           \nonumber \\
  w_{\mu} & = & \frac{1}{|\eta_{\mu+2}|^4} \left[4 \left(|\eta_{\mu+2}|^2 -
           |\eta_{\mu+1}|^2\right) + |\eta_{\mu+1}|^4 u_{\mu}\right] \qquad
\mbox{if}
           \qquad 1 + \alpha_{\mu+2} = r_{\mu+2} = 0.   \label{eq:sol2}
\end{eqnarray}
\par
%
%------------------------------------------------------------------------
%
The first solution, given in Eq.~(\ref{eq:sol1}), exists for any algebra
\algthree,
since no condition, other than Eq.~(\ref{eq:cond-Fock}), is imposed on the
parameters $\alpha_0$,~$\alpha_1$. Such a solution is of Rubakov-Spiridonov
type,
as the resulting relation~(\ref{eq:cond2-gen}) only differs from
Eq.~(\ref{eq:cond2a}) by a renormalization of the parasupercharge
operators. Let us
choose $|\eta_{\mu+2}| = \left(4 - |\eta_{\mu+1}|^2\right)^{1/2}$ so that
Eqs.~(\ref{eq:cond2a}) and~(\ref{eq:cond2-gen}) coincide, and let us fix
the overall
arbitrary phase of~$Q_{\mu}$ in such a way that $\eta_{\mu+1}$ is real and
positive (hence it belongs to the interval $(0,2)$). We then conclude that the
two-parameter family of operators
\begin{eqnarray}
  Q_{\mu}(\eta_{\mu+1}, \varphi) & = & \ap \left(\eta_{\mu+1} P_{\mu+1} +
          e^{i\varphi} \sqrt{4 - \eta_{\mu+1}^2}\, P_{\mu+2}\right),
\nonumber \\
  \H_{\mu}(\eta_{\mu+1}) & = & N + \case{1}{2} \left(2\gamma_{\mu+2} +
          r_{\mu+2} - 1\right) I + 2P_{\mu+1} + P_{\mu+2},
\end{eqnarray}
where $r_{\mu+2} = (1 + \alpha_{\mu+2}) \left(1 - \frac{1}{2} \eta_{\mu+1}^2
\right)$, $0 < \eta_{\mu+2} < 2$, and $0 \le \varphi < 2\pi$, can be
associated to
the Rubakov-Spiridonov PSSQM of order two.\par
%
%-----------------------------------------------------------------------
%
{}For a given PSSQM Hamiltonian, i.e., for a given $\eta_{\mu+1}$ value,
for instance
$\eta_{\mu+1} = \sqrt{2}$, we obtain $r_{\mu+2} = 0$, and
\begin{eqnarray}
  Q_{\mu}(\varphi) & = & \ap \sqrt{2} \left(P_{\mu+1} + e^{i\varphi}
          P_{\mu+2}\right), \nonumber \\
  \H_{\mu} & = & N + \case{1}{2} \left(2\gamma_{\mu+2} - 1\right) I +
2P_{\mu+1}
          + P_{\mu+2}.   \label{eq:sol-RS}
\end{eqnarray}
{}From the supercharge operators given in Eq.~(\ref{eq:sol-RS}), we may
single out
the two linear combinations with real coefficients
\begin{equation}
  Q_{1\mu} \equiv Q_{\mu}(0) = \ap \sqrt{2} \left(P_{\mu+1} + P_{\mu+2}\right),
  \qquad Q_{2\mu} \equiv Q_{\mu}(\pi) = \ap \sqrt{2} \left(P_{\mu+1} -
  P_{\mu+2}\right),
\end{equation}
which may be taken as the two independent conserved parasupercharges
with~$p=2$, whose existence was established by Khare~\cite{khare}. It can indeed
be checked that $Q_{1\mu}$ and~$Q_{2\mu}$ not only satisfy Eqs.~(\ref{eq:cond1})
and~(\ref{eq:cond2a}), but also the mixed trilinear equations given in
Ref.~\cite{khare}, and involving two appropriately chosen bosonic constants.\par
%
%------------------------------------------------------------------------
%
The spectra of the PSSQM Hamiltonians~$\H_{\mu}$, $\mu=0$, 1,~2, defined in
Eq.~(\ref{eq:sol-RS}), and the action of the corresponding parasupercharge
operators~$\Qp_{\mu}$ are schematically illustrated on Fig.~2. For convenience
sake, the ground states of the three spectra have been drawn on the same level,
although their energies are in general different, since they are given by
$(2\gamma_2 - 1)/2$, $(2\gamma_0 + 1)/2$, and $(2\gamma_1 + 3)/2$ for
$\mu=0$, 1, and~2, respectively. For~$\mu=0$, parasupersymmetry is unbroken,
and the ground state energy may be positive, null, or negative as $\gamma_2
> -1$
(as a consequence of Eqs.~(\ref{eq:cond-Fock}), and (\ref{eq:gamma})),
whereas for
$\mu=1$ or~2, parasupersymmetry is broken, and the ground state energy is
positive since
$\gamma_0 > - 1/2$, and $\gamma_1 > - 3/2$. We therefore recognize three of the
five possible forms of Rubakov-Spiridonov PSSQM spectra, as given in
Figs.~1 (a),
(b),~(e) of Ref.~\cite{rubakov}.\par
%
%-------------------------------------------------------------------------
%
Contrary to the first solution of Eq.~(\ref{eq:cond2-gen}), the second one,
given in
Eq.~(\ref{eq:sol2}), only exists for some algebras \algthree, namely those for
which $\alpha_{\mu+2} = -1$ (hence the value $\mu=1$ is excluded). Since the
conditions of Eq.~(\ref{eq:sol2}) imply that $v_{\mu} \ne u_{\mu}$, one may
consider the possibility of realizing Beckers-Debergh PSSQM. For
$|\eta_{\mu+2}| =
|\eta_{\mu+1}|$, and $u_{\mu} = -4 / |\eta_{\mu+1}|^2$, one indeed obtains
Eq.~(\ref{eq:cond2b}) up to a renormalization of the parasupercharge
operators. If
we choose $|\eta_{\mu+2}| = |\eta_{\mu+1}| = \sqrt{2}$ so that
Eqs.~(\ref{eq:cond2b}) and~(\ref{eq:cond2-gen}) coincide, and if we fix the
overall
arbitrary phase of~$Q_{\mu}$ in the same way as in the previous case, we recover
Eq.~(\ref{eq:sol-RS}), already obtained from Eq.~(\ref{eq:sol1}). Hence,
Equation~(\ref{eq:sol2}) does not lead to a new realization of PSSQM, but merely
shows that for those algebras for which $\alpha_{\mu+2} = -1$,
Equations~(\ref{eq:cond2a}) and~(\ref{eq:cond2b}) are simultaneously valid,
i.e., $Q
\Qp Q = Q^2 \Qp + \Qp Q^2 = 2Q\H$~\cite{beckers90}.\par
%
%========================================================================
%
\section{Conclusion}
In the present letter, we introduced $C_{\lambda}$-extended oscillator algebras
\alg, $\lambda=2$, 3,~$\ldots$, containing the Calogero-Vasiliev algebra as a
special case (corresponding to $\lambda=2$). We studied their realization as
GDOAs, and their corresponding Fock-space representation.\par
%
%------------------------------------------------------------------------
%
We then considered the bosonic oscillator Hamiltonian~$H_0$ associated to \alg,
and proved that for~$\lambda=3$, its spectrum has a very rich structure,
contrary
to what happens for~$\lambda=2$. In particular, we showed that for some
parameter values, one gets periodic spectra similar to those arising
in~SSQM with
cyclic shape invariant potentials of period three. Finally, we established that
\algthree\ provides a bosonization of Rubakov-Spiridonov PSSQM of order two. As
mentioned in previous sections, such results may be generalized to higher
$\lambda$~values.\par
%
%------------------------------------------------------------------------
%
It is clear that there remain many open problems for future study. One of them
would be a better understanding of the relationship between
$C_{\lambda}$-extended oscillator algebras and SSQM with cyclic shape invariant
potentials. Another would be an SSQM interpretation of the $H_0$~spectrum for
some of those parameter values that do not correspond to cyclic shape invariant
potentials.\par
%
%-----------------------------------------------------------------------
%
Possible connections with other extensions of SSQM than PSSQM are also worth
investigating. Realizations of pseudosupersymmetric~\cite{beckers95} and
orthosupersymmetric~\cite{mishra} quantum mechanics are under current study,
and we hope to report on them in a near future.\par
%
%========================================================================
%

%
%========================================================================
%
\newpage
\parindent0cm
\section*{Figure captions}
{\bf Fig.~1.} Energy spectra of cyclic shape invariant potentials of period
three,
obtained with Hamiltonian~$H'_0$, defined in Eq.~(\ref{eq:Hprime}): (a)
type (I.1.1)
spectrum with $\alpha_0 = 0$, $\alpha_1 = 1$; (b) type (II.1.1.1) spectrum with
$\alpha_0 = 4$, $\alpha_1 = -3$; (c) type (III.1.1.1) spectrum with
$\alpha_0 = 6$,
$\alpha_1 = -7$.

{\bf Fig.~2.} Excited energy spectra of PSSQM Hamiltonians $\H_{\mu}$, $\mu=0$,
1,~2, defined in Eq.~(\ref{eq:sol-RS}). The action of the corresponding
parasupercharge operators~$\Qp_{\mu}$ is also illustrated.
%
%=======================================================================
%
\newpage
\begin{picture}(160,120)

\savebox{\three}(10,60){\multiput(0,0)(0,30){3}{\levelone}}
\savebox{\four}(10,90){\multiput(0,0)(0,30){4}{\levelone}}
\savebox{\dash}(5,20)[b]{\multiput(0,0)(0,4){3}{\trait}}

\put(-5,-5){\thicklines\vector(0,1){115}}
\multiput(-6,0)(0,30){4}{\ord}
\put(-12,0){\makebox(5,0)[r]{\Large 0}}
\put(-12,30){\makebox(5,0)[r]{\Large $\Omega_3$}}
\put(-12,60){\makebox(5,0)[r]{\Large $2\Omega_3$}}
\put(-12,90){\makebox(5,0)[r]{\Large $3\Omega_3$}}

\multiput(50,-5)(60,0){2}{\separation}

\put(20,-15){\makebox(0,0){\Large (a)}}

\put(0,0){\usebox{\four}}
\put(5,-3){\makebox(0,0){\Large 0}}
\put(5,27){\makebox(0,0){\Large 3}}
\put(5,57){\makebox(0,0){\Large 6}}
\put(5,87){\makebox(0,0){\Large 9}}
\put(5,95){\usebox{\dash}}

\put(15,15){\usebox{\three}}
\put(20,12){\makebox(0,0){\Large 1}}
\put(20,42){\makebox(0,0){\Large 4}}
\put(20,72){\makebox(0,0){\Large 7}}
\put(20,80){\usebox{\dash}}

\put(30,25){\usebox{\three}}
\put(35,22){\makebox(0,0){\Large 2}}
\put(35,52){\makebox(0,0){\Large 5}}
\put(35,82){\makebox(0,0){\Large 8}}
\put(35,90){\usebox{\dash}}

%-----------------------------------------------------------------------

\put(80,-15){\makebox(0,0){\Large (b)}}

\put(60,0){\usebox{\four}}
\put(65,-3){\makebox(0,0){\Large 0}}
\put(65,27){\makebox(0,0){\Large 3}}
\put(65,57){\makebox(0,0){\Large 6}}
\put(65,87){\makebox(0,0){\Large 9}}
\put(65,95){\usebox{\dash}}

\put(75,5){\usebox{\three}}
\put(80,2){\makebox(0,0){\Large 2}}
\put(80,32){\makebox(0,0){\Large 5}}
\put(80,62){\makebox(0,0){\Large 8}}
\put(80,70){\usebox{\dash}}

\put(90,15){\usebox{\three}}
\put(95,12){\makebox(0,0){\Large 1}}
\put(95,42){\makebox(0,0){\Large 4}}
\put(95,72){\makebox(0,0){\Large 7}}
\put(95,80){\usebox{\dash}}

%------------------------------------------------------------------------

\put(140,-15){\makebox(0,0){\Large (c)}}

\put(120,0){\usebox{\four}}
\put(125,-3){\makebox(0,0){\Large 2}}
\put(125,27){\makebox(0,0){\Large 5}}
\put(125,57){\makebox(0,0){\Large 8}}
\put(125,87){\makebox(0,0){\Large 11}}
\put(125,95){\usebox{\dash}}

\put(135,15){\usebox{\three}}
\put(140,12){\makebox(0,0){\Large 0}}
\put(140,42){\makebox(0,0){\Large 3}}
\put(140,72){\makebox(0,0){\Large 6}}
\put(140,80){\usebox{\dash}}

\put(150,20){\usebox{\three}}
\put(155,17){\makebox(0,0){\Large 1}}
\put(155,47){\makebox(0,0){\Large 4}}
\put(155,77){\makebox(0,0){\Large 7}}
\put(155,85){\usebox{\dash}}

\end{picture}

\vspace{5cm}
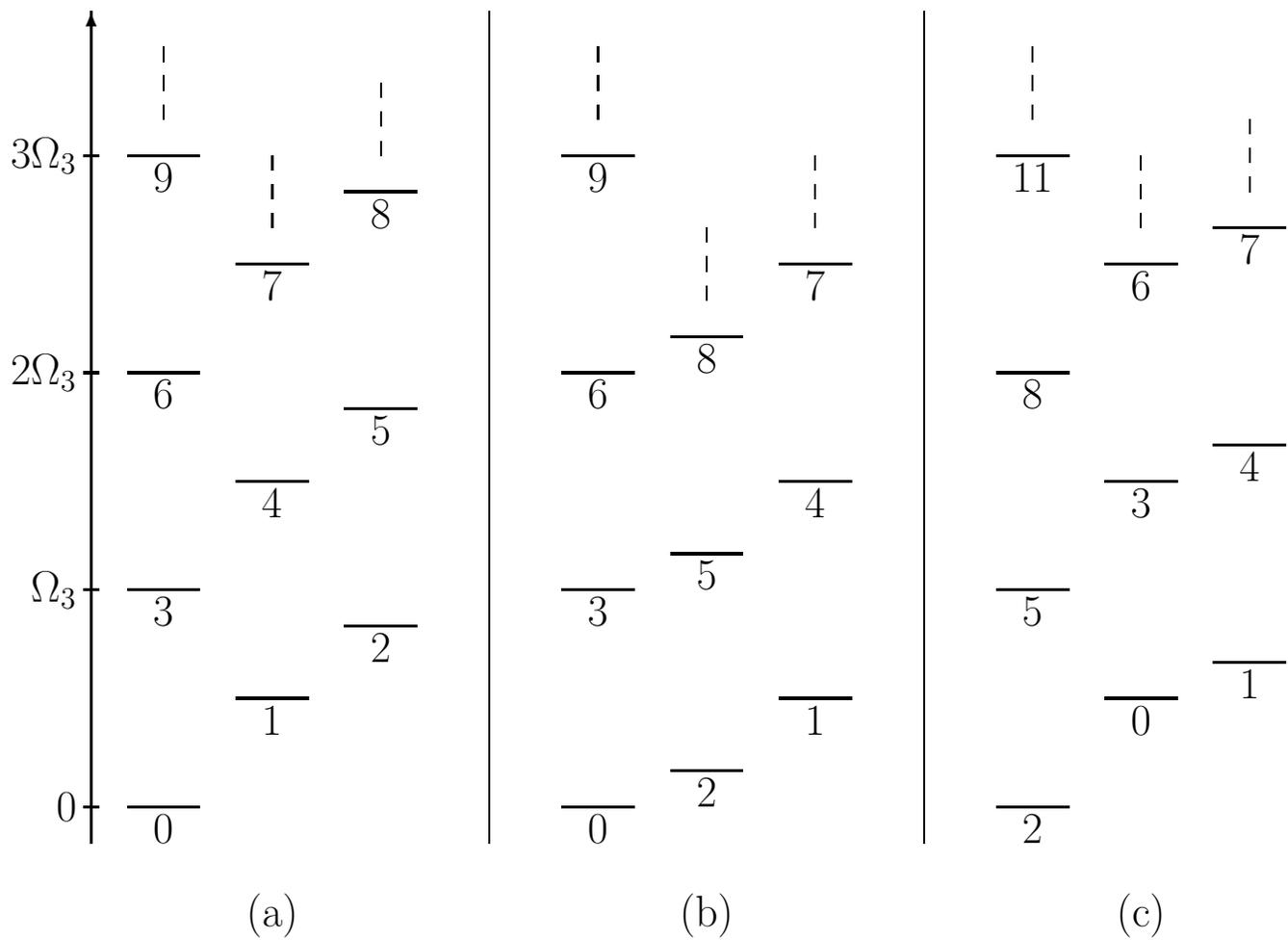
\centerline{Figure 1}
%
%========================================================================
%
\newpage
\begin{picture}(160,120)(-11,0)

\savebox{\three}(8,60){\multiput(0,0)(0,30){3}{\leveltwo}}
\savebox{\four}(8,90){\multiput(0,0)(0,30){4}{\leveltwo}}
\savebox{\dash}(5,20)[b]{\multiput(0,0)(0,4){3}{\trait}}
\savebox{\Qzero}(0,0){\Large $Q_0^{\dagger}$}
\savebox{\Qone}(0,0){\Large $Q_1^{\dagger}$}
\savebox{\Qtwo}(0,0){\Large $Q_2^{\dagger}$}

\put(-5,-5){\thicklines\vector(0,1){115}}
\multiput(-6,0)(0,30){4}{\ord}
\put(-12,0){\makebox(5,0)[r]{\Large 0}}
\put(-12,30){\makebox(5,0)[r]{\Large 3}}
\put(-12,60){\makebox(5,0)[r]{\Large 6}}
\put(-12,90){\makebox(5,0)[r]{\Large 9}}

\multiput(45,-5)(50,0){2}{\separation}

\put(20,-15){\makebox(0,0){\Large $\mu=0$}}

\put(0,0){\usebox{\four}}
\put(10,0){\makebox(0,0){\Large 0}}
\put(10,30){\makebox(0,0){\Large 3}}
\put(10,60){\makebox(0,0){\Large 6}}
\put(10,90){\makebox(0,0){\Large 9}}
\put(4,95){\usebox{\dash}}

\put(16,30){\usebox{\three}}
\put(26,30){\makebox(0,0){\Large 2}}
\put(26,60){\makebox(0,0){\Large 5}}
\put(26,90){\makebox(0,0){\Large 8}}
\put(20,95){\usebox{\dash}}

\put(32,30){\usebox{\three}}
\put(42,30){\makebox(0,0){\Large 1}}
\put(42,60){\makebox(0,0){\Large 4}}
\put(42,90){\makebox(0,0){\Large 7}}
\put(36,95){\usebox{\dash}}

\qbezier(4,30)(12,50)(20,30)
\put(13,40){\thicklines\vector(2,0){0}}
\put(12,45){\usebox{\Qzero}}

\qbezier(20,30)(28,50)(36,30)
\put(29,40){\thicklines\vector(2,0){0}}
\put(28,45){\usebox{\Qzero}}

%-----------------------------------------------------------------------

\put(70,-15){\makebox(0,0){\Large $\mu=1$}}

\put(50,0){\usebox{\four}}
\put(60,0){\makebox(0,0){\Large 1}}
\put(60,30){\makebox(0,0){\Large 4}}
\put(60,60){\makebox(0,0){\Large 7}}
\put(61,90){\makebox(0,0){\Large 10}}
\put(54,95){\usebox{\dash}}

\put(66,0){\usebox{\four}}
\put(76,0){\makebox(0,0){\Large 0}}
\put(76,30){\makebox(0,0){\Large 3}}
\put(76,60){\makebox(0,0){\Large 6}}
\put(76,90){\makebox(0,0){\Large 9}}
\put(70,95){\usebox{\dash}}

\put(82,30){\usebox{\three}}
\put(92,30){\makebox(0,0){\Large 2}}
\put(92,60){\makebox(0,0){\Large 5}}
\put(92,90){\makebox(0,0){\Large 8}}
\put(86,95){\usebox{\dash}}

\qbezier(54,30)(62,50)(70,30)
\put(63,40){\thicklines\vector(2,0){0}}
\put(62,45){\usebox{\Qone}}

\qbezier(70,30)(78,50)(86,30)
\put(79,40){\thicklines\vector(2,0){0}}
\put(78,45){\usebox{\Qone}}

%-----------------------------------------------------------------------

\put(120,-15){\makebox(0,0){\Large $\mu=2$}}

\put(100,0){\usebox{\four}}
\put(110,0){\makebox(0,0){\Large 2}}
\put(110,30){\makebox(0,0){\Large 5}}
\put(110,60){\makebox(0,0){\Large 8}}
\put(111,90){\makebox(0,0){\Large 11}}
\put(104,95){\usebox{\dash}}

\put(116,0){\usebox{\four}}
\put(126,0){\makebox(0,0){\Large 1}}
\put(126,30){\makebox(0,0){\Large 4}}
\put(126,60){\makebox(0,0){\Large 7}}
\put(127,90){\makebox(0,0){\Large 10}}
\put(120,95){\usebox{\dash}}

\put(132,0){\usebox{\four}}
\put(142,0){\makebox(0,0){\Large 0}}
\put(142,30){\makebox(0,0){\Large 3}}
\put(142,60){\makebox(0,0){\Large 6}}
\put(142,90){\makebox(0,0){\Large 9}}
\put(136,95){\usebox{\dash}}

\qbezier(104,30)(112,50)(120,30)
\put(113,40){\thicklines\vector(2,0){0}}
\put(112,45){\usebox{\Qtwo}}

\qbezier(120,30)(128,50)(136,30)
\put(129,40){\thicklines\vector(2,0){0}}
\put(128,45){\usebox{\Qtwo}}

\end{picture}

\vspace{5cm}
\centerline{Figure 2}

\end{document}